\documentclass[aps,twocolumn]{revtex4}

\usepackage{epsfig}
\usepackage{graphics}
\usepackage{color}

\begin{document}
\title{Interaction induced phase fluctuations in a guided atom laser}
\author{Isabelle Bouchoule$^{(1)}$ and Klaus M{\o}lmer$^{(2)}$}
\affiliation{(1) : Institut d'Optique, 91 403 ORSAY Cedex, France\\
(2) : QUANTOP -- Danish National Reserach Foundation Center for Quntum
Optics, Department of Physics and Astronomy, University of Aarhus, 
DK 8000 Aarhus C., Denmark
}

\begin{abstract}
In this paper, we determine the magnitude of phase fluctuations 
caused by atom-atom interaction in a one-dimensional beam of bosonic 
atoms.  We imagine that the beam is created with a large
coherence length, and that interactions only act
in a specific section of the beam, where the atomic density is high
enough to validate a Bogoliubov treatment.
The magnitude and coherence length of the ensuing phase fluctuations 
in the beam after the interaction zone are determined.
\end{abstract}


\maketitle

Recent progress on micro-fabricated magnetic traps and guides for
atoms \cite{JAcobBEC,BECZimmermann}  have stimulated the efforts to
realize guided atom-laser beams \cite{BECmicroKetterle} with
possible applications for high precision atom interferometry.
The phase coherence is a crucial property 
of such beams and interferometers, and interactions
between atoms may influence the coherence. So far, the effect of 
interactions on the condensate phase were mainly studied in 
condensates of finite spatial extent \cite{Lewenstein}, where both 
degradation \cite{JavaWil} and squeezing of the phase 
\cite{Uffe} have been studied. In these cited works, the main effect 
can be ascribed to the number fluctuations in the condensate and the 
dependence of the mean field energy on atom number due to interactions. 
The goal of the present work is to determine the amplitude and the 
coherence length of the phase fluctuations created by the atomic 
interactions in a continuous beam whose coherence length is initially 
infinite.
The main contribution in this case is 
collisions that transfer atoms from the condensate at momentum $K$
to different momenta $K+k$ in the beam. 
At very low density an interacting bose gas in 
one dimension is described by the Tonks-Girardeau regime, but
we shall assume  a higher density for which Bogoliubov theory 
is well suited to describe the deviations from 
mean field theory \cite{TFTonk_olsh01}.

We consider the  situation depicted in Fig.1, 
where a non-interacting mono-energetic beam of atoms with momentum $\hbar K$ 
arrives in a region of length $L$ where the atoms interact with each other. 
For the practical realization of such a model, one may 
imagine a guided atom beam, where the
transverse width is particular narrow, and therefore the density and the
effective interaction is significantly increased over the length $L$.

\begin{figure}[h]
\centerline{\includegraphics{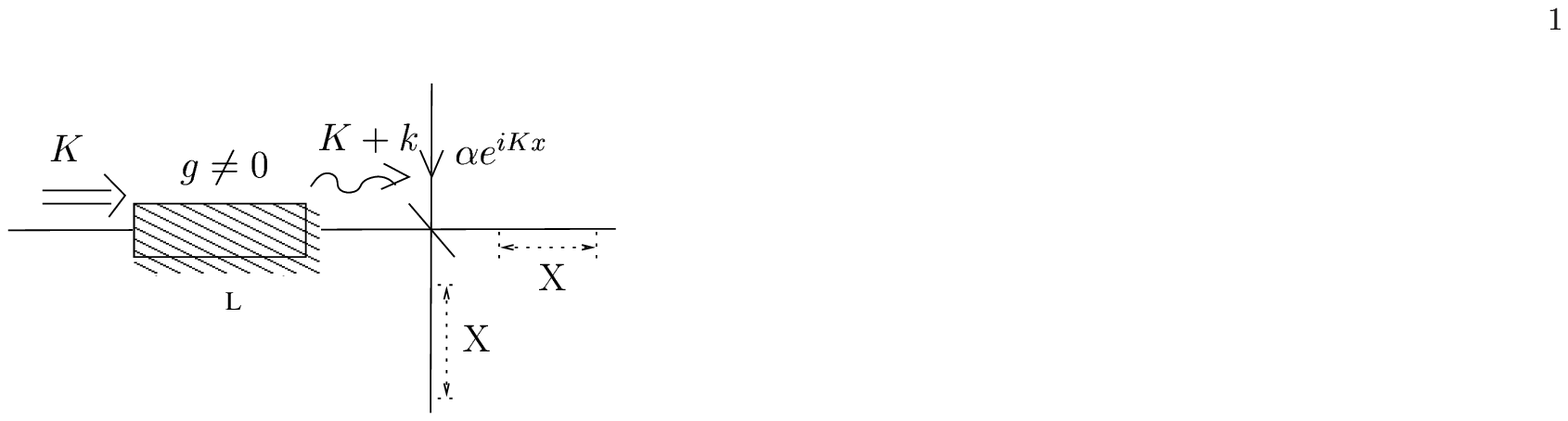}}
\caption{An atomic beam crosses a region $L$ where the atoms interact
with each other. The interactions scatter atoms out of the macroscopically
populated mode into other momentum states, and the resulting phase 
fluctuations in the beam can be probed by the interference with a 
coherent reference beam.}
\end{figure}

The system is described by the effective one dimensional Hamiltonian 
\begin{equation}
H\!=\!\! \int \!\! dx 
\psi^\dagger(x) h_0 \psi(x)
+\frac{g}{2}\int_0^L\!\! dx \psi^\dagger(x)\psi^\dagger(x)\psi(x)\psi(x),
\label{eq.Ham2quant}
\end{equation}
where $h_0=-\hbar^2/(2m)d^2/dx^2$, and the coupling constant
$g=\frac{2\hbar^2}{m}\frac{2a}{a_{\perp}^2}$,
assuming that the size of the transverse ground state   
in the guide $a_\perp=\sqrt{2\hbar/m\omega}$
is much larger than the s-wave scattering length $a$.

We always consider the case where $n a_{\perp}^2/a >> 1$
so that correlations between atoms are small,
and the state of the system can be described by a mean field
wave function and a noise term
\begin{equation}
\psi(x)=\sqrt{n} e^{iKx} +\delta \psi(x).
\label{fieldop}
\end{equation}
The fluctuations represented by $\delta \psi(x)$ are probed 
by interference with a reference beam in a coherent state 
$\psi_{ref}(x)=\alpha e^{iKx}$.
The quantity measured is the number of atoms on each side of the 
beam splitter. In an actual experiment, this beam 
splitter may be realized by tunneling  between 
two guides, or the atomic  beams may be 
physically overlapping, but with
two different internal states, which can be coupled by Raman laser
beams. In either case,
the operators giving the density of atoms in the two output states are
\begin{equation}
n_{\pm}(x)=\frac{1}{2}(
\psi_{ref}^{\dagger}(x) \pm \psi^{\dagger}(x) ) 
\left ( \psi_{ref}(x)\pm \psi(x) \right )
\end{equation}
For $\alpha=\sqrt{n_0} e^{i\pi/2}$, the mean density is identical in the
two output beams and we define the local phase operator as 
\begin{equation}
\theta(x)=\frac{ n_+(x)-n_-(x)}{\sqrt{n_0 n}}. 
\end{equation} 
The fluctuations in $\theta(x)$ are given by the correlation 
function
\begin{equation}
\langle \theta (x) \theta (x') \rangle =
\frac{1}{n}
\begin{array}[t]{l}
\{
- \langle \delta\psi(x')\delta\psi(x)\rangle e^{-iK(x+x')}
-c.c.  \\
+\langle \delta\psi^{\dagger}(x')\delta\psi(x) \rangle e^{iK(x'-x)}
+ c.c. \\
 +\delta(x'-x)\},
\end{array}
\label{eq.fluctuphase}
\end{equation}
and it is our goal to present a quantitative analysis
of these fluctuations.
After averaging over a distance $X$, the fluctuations of 
the phase is given by
\begin{equation}
\Delta \theta^2=\left \langle  \left ( 
\frac{1}{X} \int_X \theta (x) dx \right )^2 \right \rangle,
\end{equation}
In the absence of interactions between atoms, only the  last
$\delta$-correlated quantum noise term is present in
Eq.(\ref{eq.fluctuphase}), and we find
$\Delta \theta=1/\sqrt{nX}$ 
as expected from the usual number and phase uncertainties in a coherent 
state. 

In the presence of interactions between atoms, a condensate 
will experience phase diffusion which is due to 
the spread of chemical potential over the Poisson distribution of number 
states which form the coherent state.
Let us assume that the atomic beam is in fact a very long wavepacket
with length $\cal L$. When a superposition of number states $|N\rangle$
of this wavepacket passes the interaction zone the interaction perturbs
the energy of each number state by the amount
$\frac{g}{2}n^2L=\frac{g}{2}\frac{N^2}{{\cal L}^2}L$. The passage time
is $T={\cal L}/(\hbar K/m)$ and hence the number state component
$|N\rangle$ experiences a phase shift $\theta_N=\frac{gmL}{2{\cal L}\hbar^2
K} N^2$. 
The phase of a coherent state is the derivative $d\theta/dN$ of the phase 
of the Fock components. After passage of the interacting zone,  
this derivative depends on $N$ and the spread of 
the phase over the width $\sqrt{N}$ of the 
Fock state distribution is 
\begin{equation}
\Delta \theta =
\sqrt{N}\frac{gmL}{{\cal L}\hbar^2 K}=
\frac{\sqrt{n}gmL}{\sqrt{{\cal L}}\hbar^2 K}. 
\label{Goldstone}
\end{equation}
The relative number fluctuations vanish in the
limit of very large coherence length $\cal L$ of the incident beam, 
and in that limit the Bogoliubov 
excitations will give the dominant contribution to phase diffusion.
 
If the expression for the field operator (\ref{fieldop}) is introduced
in the second quantized Hamiltonian (\ref{eq.Ham2quant}), and 
the noise terms $\delta\psi(x),\delta\psi^\dagger(x)$ are 
truncated above second order, $H$ can be diagonalized
by a Bogoliubov transformation,
$H=E_0 + \sum \omega_k \alpha^\dagger_k \alpha_k$,
where the field operator $\delta\psi(x)$ is
expanded on the bosonic operators
$\alpha_k,\alpha_k^\dagger$ as
$\delta\psi(x)=\sum_k \left 
( u_k(x)\alpha_k -v_k^*(x)\alpha_k^\dagger \right )$.
The wavefunctions $u_k$ and $v_k$ solve the equations
\begin{equation}
\left ( 
\begin{array}{cc}
h_0 -\mu +2g|\varphi|^2 & -g\varphi^2\\
g\varphi^{*2}&-h_0 +\mu -2g|\varphi|^2
\end{array}
\right )
\left (
\begin{array}{c}
u_k\\
v_k
\end{array}
\right )
=
\omega_k 
\left (
\begin{array}{c}
u_k\\
v_k
\end{array}
\right ),
\label{eq.Bogo}
\end{equation}
where $g$ takes the value zero outside the interaction interval,
and $u_k$ and $v_k$ fulfill the normalization condition
$\int dx (u_k(x)u_{k'}(x)^*-v_k(x)v_{k'}(x)^*) =\delta(k-k')$.

We neglect reflection of the mean field wave
function at the entrance and exit of the
interaction region, which is a good approximation as long
as $gn \ll \hbar^2 K^2/2m$. When we write 
the mean field wave function in the interaction zone as 
$\varphi(x)=\sqrt{n}e^{iKx}$,
the chemical potential is $\mu=\hbar^2K^2/2m+gn$,
and Eq.(\ref{eq.Bogo}) has the plane wave solutions 
\begin{equation}
\left (
\begin{array}{c}
u\\
v
\end{array}
\right )=
\left (
\begin{array}{c}
e^{i(k+K)x}U\\
e^{i(k-K)x}V
\end{array}
\right ).
\end{equation} 
Inserting this expansion  in Eq.(\ref{eq.Bogo}) we obtain
\begin{equation}
\left \{
\begin{array}{l}
\left ( \frac{\hbar^2}{2m}(k+K)^2 
-\mu +2 gn  \right )U -gn V=\omega U\\
\left ( \frac{\hbar^2}{2m}(k-K)^2 
-\mu +2gn  \right )V -gn U=-\omega V\\
\end{array}
\right .
\label{eq.bogohom}
\end{equation}
Outside the interaction zone, where $g=0$, $U$ and $V$ are independent with
energies $\omega_{u/v}=kK\pm k^2/2$. 
Inside the interaction zone where  $g\neq 0$, we multiply the first 
equation of Eq.(\ref{eq.bogohom}) by $\left ( \frac{\hbar^2}{2m}(k-K)^2 -\mu +2
gn  \right )$
and obtain a second order equation for $\omega$ which leads to 
\begin{equation}
\omega=\frac{\hbar^2}{m}kK\pm 
\sqrt{\frac{\hbar^2 k^2}{2m}\left (\frac{\hbar^2k^2}{2m}+2 gn \right )}
\label{eq.wk}
\end{equation}
The spectrum has  two branches corresponding for high energies to 
particle and hole-like excitations, respectively.
For the two branches, the ratios $V/U$ of the solutions are
\begin{equation}
\left \{
\begin{array}{l}
y_1=\frac{V_1}{U_1}
=1+\frac{\hbar^2}{m gn}
\left (\frac{k_1^2}{2}+\sqrt{\frac{k_1^2}{2}(\frac{k_1^2}{2}+2m gn/\hbar^2)}
\right )\\
y_2=\frac{V_2}{U_2}
=1+\frac{\hbar^2}{m gn}
\left (\frac{k_2^2}{2}-\sqrt{\frac{k_2^2}{2}(\frac{k_2^2}{2}+2m gn/\hbar^2)}
\right )\\
\end{array}
\right .
\label{eq.y}
\end{equation} 

In each region along the atomic beam, four solutions are possible, 
corresponding in the noninteracting region to particles or holes propagating in 
one direction or the other, and the global solutions should be
continuous with continuous derivatives
for both the $u$ and $v$ functions.
Thus 8 equations relate the 12 parameters and for each energy 
4 solutions exist.
 Like in normal scattering theory,
our choice of boundary conditions serves to identify the 
relevant basis for the solution to the problem.
Excitations with an escaping hole (removal of an incident particle) 
in either direction are not physically relevant, and we thus study 
the elementary excitations with incoming particle components only.

Let us first consider the excitation corresponding to an incoming 
particle $u(x) \propto e^{i(K+k)x}$ from the right, ($k<-K$).
This particle has a momentum which differs from that of the condensate
by more than $K$. If we assume that rather than being an exact
delta-function, the atomic interaction has a momentum cut off which is
smaller than $K$, 
the coupling to the $v(x)$ function in the
Bogoliubov equations is suppressed, and 
this mode does  not contribute to phase fluctuations
in the Bogoliubov vacuum. 
Another set of Bogoliubov modes 
corresponds to incoming particles from the left with momentum $k+K$ 
and amplitude $U_i=\frac{1}{\sqrt{2\pi}}$ 
(the condition 
$[\delta\psi(x),\delta\psi^\dagger(x')]=\delta(x-x')$ serves to
normalize the Bogolibuv mode since
$\int_k dk   U_i(k)e^{i(K+k)x}U_i(k)^*e^{-i(K+k)x'}=\delta(x-x')$
on the left side where the Bogoliubov modes contain no holes).
Eq.(\ref{eq.wk}) has  four solutions, but 
we will consider only momentum components $K+k_1$ and $K+k_2$ close to 
$K$ inside the interaction zone, since the effective momentum 
cut-off in the interaction prevents coupling to waves with very different 
momenta. Continuity of the solution at the left entrance to the 
interaction zone implies
\begin{equation}
\left \{
\begin{array}{l}
U_i=U_1+U_2\\
0=V_1+V_2
\end{array}
\right . 
\end{equation}
Eq.(\ref{eq.y}) dictates the ratio between the $U$ and $V$ amplitudes, 
and the continuity of the derivatives of $u$ and $v$ cannot be fulfilled 
without allowing for reflection.
This reflection will, however, be very small as $k\ll K$, and we make
an insignificant error by requiring only continuity of the
Bogoliubov mode functions.  All amplitudes in the inteaction zone are
hence given by the incident amplitude $U_i=\frac{1}{\sqrt{2\pi}}$,
\begin{equation}
\left \{
\begin{array}{l}
V_1=U_i\frac{y_1y_2}{y_2-y_1}\\
V_2=U_i\frac{y_1y_2}{y_1-y_2}\\
\end{array}
\right . \mbox{ and }
\left \{
\begin{array}{l}
U_1=U_i\frac{y_2}{y_2-y_1}\\
U_2=U_i\frac{y_1}{y_1-y_2}\\
\end{array}
\right . 
\label{eq.contAB}
\end{equation}
At the right of the interaction zone, we have to match to the
independent solutions $Ue^{i(K+k)x}$ and $Ve^{-i(K-k')x}$ of same
Bogoliubov energy $\omega$, i.e., the amplitudes $U$ and $V$ 
are given by the continuity relations
\begin{equation}
\left \{
\begin{array}{l}
Ue^{i(k+K)L}=U_1e^{i(k_1+K)L}+U_2e^{i(k_2+K)L}\\
Ve^{i(k'-K)L}=V_1e^{i(k_1-K)L}+V_2e^{i(k_2-K)L}\\
\end{array}
\right .
\label{eq.contBC}
\end{equation}
Using the relations (\ref{eq.contAB}), this gives
\begin{equation}
\left \{
\begin{array}{l}
U=U_i e^{-ikL}\left ( \frac{y_2}{y_2-y_1} e^{ik_1L}
+\frac{y_1}{y_1-y_2} e^{ik_2L} \right )\\
V=U_i e^{-ik'L}\left ( \frac{y_1 y_2}{y_1-y_2} \right )
e^{i\frac{k_1+k_2}{2}L}2i\sin\left ( \frac{k_1-k_2}{2}L\right )
\end{array}
\right .
\label{eq.vu}
\end{equation}

The phase fluctuations given 
in Eq.(\ref{eq.fluctuphase}) can now be expressed as
\begin{equation}
\langle \theta(x)\theta(x') \rangle=
\frac{1}{n}\int dk
\begin{array}[t]{l}
\{ 2{\cal R}e ( V(k)^* U(k) e^{ikx'}e^{-ik'x})  \\
 + 2|V(k)|^2 \cos(k'(x'-x)) \},
\end{array}
\label{eq.intphase}
\end{equation}
where the $\delta$-function part has been omitted.

We now present a number of simplifications, which are valid under 
reasonable assumptions and which will lead to a better understanding 
of the dependence of the phase fluctuations on the physical parameters 
of the problem.  First, as long as 
$k_1 \simeq k_2 \simeq k \ll K$ we can use the expressions
of $y_1$ and $y_2$ (\ref{eq.y}) for the single value $k$.
We cannot, however, replace $k_1$ and $k_2$ generally by $k$ in the
exponential factors in (\ref{eq.vu}). Instead, we look at limiting
cases.  If $k\gg \sqrt{m gn}/\hbar$, the energy $\omega=kK+k^2/2$ leads
to the same dispersion law as that of a non interacting gas. 
This dispersion law leads to the expressions, $k_1\simeq k+\frac{k^2}{K}$ and
$k_2\simeq k$, and the expressions for $y_1$ and $y_2$ are in this limit 
approximated by $y_1\simeq \frac{\hbar^2 k^2}{m gn}$ and 
$y_2\simeq \frac{m gn}{\hbar^2 k^2}$.
The suppression of the $V(k)$ amplitude in Eq.(\ref{eq.vu}) by the factor
$\frac{y_1y_2}{y_1-y_2} \simeq \frac{mgn}{\hbar^2k^2}\ll 1$ results in only
negligible contributions to phase fluctuations from these momentum
components.  
If  instead, $|k| \ll \sqrt{mgn}/\hbar$,
the Bogoliubov dispersion law approximately yields
$k_1\simeq k+k\frac{\sqrt{m gn}}{\hbar K}$ and 
$k_2\simeq k-k\frac{\sqrt{m gn}}{\hbar K}$,
and the expressions for $y_1$ and $y_2$ are in this limit
approximated by $ y_1\simeq 1+\frac{\hbar |k|}{\sqrt{m gn}}$ and 
$y_2\simeq 1-\frac{\hbar |k|}{\sqrt{m gn}}$.
We then find,
\begin{equation}
U(k)
\begin{array}[t]{l}
\simeq V(k)\\
\simeq 
-\frac{e^{ikL}}{\sqrt{2\pi}}2i\frac{\sqrt{m gn}}{2\hbar k}
e^{i\frac{k_1+k_2}{2}L}
\sin\left ( \frac{k\sqrt{m gn}}{\hbar K}L\right ).
\end{array}
\label{eq.analUV}
\end{equation}
In Eq.(\ref{eq.intphase}) $k'$ is the momentum associated with the
$v$-component in the noninteracting region and,
as long as $|k|\ll K$, it writes $k'\simeq k + \frac{k^2}{K}$, and the 
cross term in $V(k)^*U(k)$ can thus be written, 
\begin{equation}
I(x,x')=\frac{2}{n}\int dk |V(k)|^2 {\cal R}e(e^{ik(x'-x)}e^{ixk^2/K}).
\label{Iuv}
\end{equation}
Approximating $k'$ by $k$ the $|V(k)|^2$
term in Eq.(\ref{eq.intphase}) can be written
\begin{equation}
J(x,x')=
\frac{2}{n} \int dk  |V(k)|^2 \cos(k(x-x')).
\label{Jvv}
\end{equation}
For large $x$, $I(x,x')$ becomes wider as a function of $y=(x'-x)$.
If we label $x/K$ as $t/2m$ in the last exponential we recover the 
expression for the spreading wave packet of a massive particle, 
given by the momentum spread $\Delta k$ of $|V(k)|^2$, i.e., 
$\Delta y = \hbar \Delta k\frac{x}{K}=\frac{\hbar x}{L\sqrt{m gn}}$.
At distances far from the interacting region $I(x,x')$
gives thus a negligible contribution to local
phase fluctuations, but when integrated  over intervals larger than
$\Delta y$, the conserved 'norm' of the spreading wave packet implies 
a result comparable to the contribution from the
$|V(k)|^2$ term, estimated below.

Fig. 2a). shows the value of $\langle \theta(x)\theta(x')\rangle$ for
different values of $x$ and as functions of $x'$. The curves are obtained
by a numerical integration of Eq.(\ref{eq.intphase}) with the proper 
expressions for $U(k)$ and $V(k)$. For large $x$, we both see 
the spreading $I(x,x')$ and the narrow $J(x,x')$ component, cf. Fig 2b).

We are now in position to obtain the essential scaling of the phase 
fluctuations with the physical parameters of the problem:
If we focus on the narrow part $J(x,x')$,
at distances from the interaction zone larger than $L^2 gn/K$,
the phase fluctuation have a finite coherence length,
estimated as the inverse of the width in momentum space of $V(k)$,
\begin{equation}
\Delta x \simeq \frac{\sqrt{mgn}L}{\hbar K}.
\label{deltax}
\end{equation}
The dependence on the size of the interacting region $L$ 
is due to a phase matching 
conditions : momentum and energy conservation cannot be both fullfiled 
in a collision which creates excitations and the violation of momentum 
conservation becomes more severe as $L$ increases. 
The amplitude of the fluctuations increases linearly with $L$,
\begin{equation}
\left \langle \theta(x)^2 \right \rangle =\frac{1}{n} 
\left (\frac{m gn}{\hbar^2}    \right )^{3/2}\frac{L}{K},
\label{local}
\end{equation} 
and to recover a well defined phase, one has to integrate over a finite
detector region in space.
If one detects atoms on a length $X$ larger than 
$\Delta x$, the phase precision is given by the $k=0$ component 
of the fourier transform of the phase fluctuations, 
\begin{equation}
\Delta \theta^2=\frac{4\pi}{n}|V(0)|^2 = \frac{2}{nX}
\left (\frac{mgnL}{\hbar^2 K} \right )^2
= 2\frac{\left \langle \theta(x)^2 \right \rangle}{X/\Delta x}.
\label{unc}
\end{equation}
If $X$ becomes  larger than the length scale $\Delta y$, the 
integral of $J(x,x')$ contributes with the same
amount, i.e., Eq.(\ref{unc}) is multiplied by 
a factor of 2.  This analysis is confirmed in Fig. 2c)., showing as
a dashed line the analytical result (\ref{unc}) and as a solid line the 
results of the full numerical calculation. The dot-dashed line in the
figure shows the value (22), applicable for short intervals $X<\Delta
x$.

Let us conclude with a numerical example for
a beam of Rubidium atoms with velocity $\hbar K/m=80\,$mm/s,  
and a linear density of $n=10^6$ atoms/m, subject to
an interacting region with a perpendicular confinement 
of $a_\perp=0.5\,\mu$m  and a length of $L=1\,$cm. 
With the value $a=5\,$nm for the scattering length we 
then obtain 
\begin{equation}
\Delta x \simeq 26\,\mu\rm{m}
\;\; 
\rm{and}
\;\;
\left \langle \theta(x)^2 \right \rangle\simeq 2.
\end{equation}
To have a phase uncertainty $\Delta \theta$ as small as $0.1$, 
one has to count the Rb atoms in this example over a length of order 
2 $cm$. Note that the phase uncertainty is two orders of magnitude
larger than the standard result for a perfect coherent state on that
interval, and if we assume a coherence length ${\cal L}$ of the beam 
exceeding 2 cm,, the contribution from number fluctuations (7) is
smaller than the Bogoliubov excitation result by at least a factor of
two.

To summarize, we have quantified the phase fluctuations in an atomic beam, 
and found that they manifest themselves at different observational length 
scales:
If $X\sim 1/n$, the number of atoms observed is of order
unity and the phase is always ill-defined.
The phase uncertainty falls off when $X$ is
increased, but in the presence of interactions, until  
$X$ reaches the length scale $\Delta x= \sqrt{mgn}L/\hbar K$, 
it levels off to a value
given by Eq.(\ref{local}). For $X$ larger 
than $\Delta x$, it decreases according to Eq.(\ref{unc}) or 
Eq.(\ref{unc}) multiplied by a factor two if $X$ is also larger than
$\Delta y=\hbar x_0/(L\sqrt{mgn})$, where $x_0$ is the 
location of the detection interval. If the coherence length 
${\cal L}$ of the beam is short, there is an extra contribution 
(\ref{Goldstone}) to phase fluctuations. Finally, it should be
mentioned that other sources of phase fluctuations may be 
important in a guided atomic beam. For example, the decoherence
and heating of the atomic beam due to the presence
of the surfaces of the macroscopic elements that provide the
guiding potential has been analysed in \cite{Henkel}.

\clearpage

\vspace*{1cm}
\begin{figure}[h]
\includegraphics{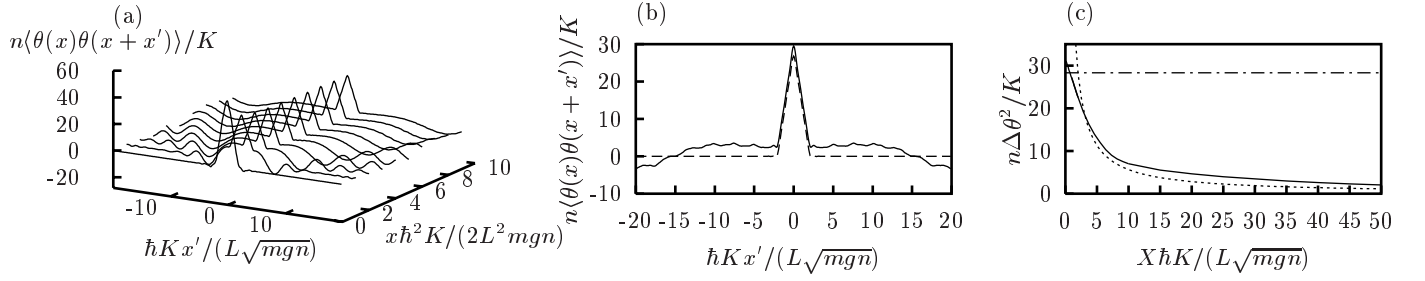}
\parbox{18cm}{\caption{\it Phase fluctuations calculated for an exanmple with 
numerical parameters
$L=10^4/K$ and $gn=0.02 \hbar^2K^2/m$.
(a) : Phase fluctuations given by Eq.(\ref{eq.intphase}).
 For large $x$ the curves clearly separate in a central
feature and a broad pedestal due to the $|V(k)|^2$ and the $V(k)^*U(k)$
contributions, respectively.
(b) : Phase correlations around $x_0=20L^2gn/K$. The $J(x,x')$ term
is shown as a dashed line.
(c): Phase fluctuations averaged over intervals of variable length $X$.
The curves show $\langle (\int_{x_0}^{x_0+X} dx \theta(x)/X)^2\rangle$
for $x_0=50 L^2gn/K$. The full line is the result of a numerical 
calculation, the dashed line is the analytical expression (\ref{unc}), and
the horizontal dot-dashed line is the value (\ref{local}) expected for 
$X$ smaller than the coherence length $\Delta x=\sqrt{gnm}L/\hbar K$.
}}
\end{figure}

\end{document}